\documentclass[twoside,twocolumn]{article}

\usepackage{blindtext} 

\usepackage[sc]{mathpazo} 
\usepackage[T1]{fontenc} 
\usepackage{microtype} 
\usepackage{hyperref}
\hypersetup{colorlinks,urlcolor=blue,linkcolor=black,citecolor=black}
\usepackage[english]{babel} 
\usepackage{paralist}
\usepackage{url}
\usepackage[hmarginratio=1:1,top=32mm,columnsep=20pt]{geometry} 
\usepackage[small,labelfont=bf,up,textfont=up]{caption} 
\usepackage{booktabs} 

\usepackage{paralist}
\usepackage{xcolor}
\usepackage{graphicx}
\usepackage{subfig}
\usepackage{chngcntr}
\usepackage{algorithm}
\usepackage[noend]{algpseudocode}
\usepackage{ amssymb }

\usepackage{microtype}

\usepackage{lettrine} 

\usepackage{enumitem} 
\setlist[itemize]{noitemsep} 

\usepackage{abstract} 

\usepackage{titlesec} 
\titleformat{\section}[block]{\large\scshape\centering}{\thesection.}{1em}{} 
\titleformat{\subsection}[block]{\large}{\thesubsection.}{1em}{} 

\usepackage{fancyhdr} 
\usepackage{titling} 

\usepackage{hyperref} 
\usepackage{authblk}

\pretitle{\begin{center}\LARGE\bfseries} 
\posttitle{\end{center}} 
\title{Time Is of the Essence: Analyzing the Effect of Vertex-Joining Time on Complex Network Evolution} 

\author[1,2]{Michael Fire}
\author[1]{Carlos Guestrin}
\affil[1]{Department of Computer Science \& Engineering, University of Washington}
\affil[2]{The eScience Institute, University of Washington}

\date{\today} 
\renewcommand{\maketitlehookd}{%
\begin{abstract}
Complex networks have non-trivial characteristics and appear in many real-world systems.
 Such networks are vitally important, but their full underlying dynamics are not completely understood. 
 Utilizing new data sources, however, can unveil the evolution process of these networks.

This study uses the recently published Reddit dataset, containing over 1.65 billion 
comments, to construct the largest publicly available social network corpus to date. 
We used this dataset to deeply examine the network evolution process, 
which resulted in two key observations: 
First, links are more likely to be created among users who join a network at a similar time. 
Second, the rate in which new users join a network is a central factor in molding a network's topology;
 i.e., different user-join patterns create different topological properties.  

Based on these observations, 
we developed the \textit{Temporal Preferential Attachment} random network generation model. 
This model produces not only scale-free networks that have relative high clustering coefficients, 
but also networks that are sensitive to both the rate and the time in which users join the network. 
This results in a more accurate and flexible model of how complex networks evolve, one which more closely represents real-world data. 
\\ \\
\textbf{Project Links:} 
 $\blacktriangleright$\href{http://homes.cs.washington.edu/~fire/reddatait/}{ \underline{Website} }
 $ \blacktriangleright$ \href{http://homes.cs.washington.edu/~fire/reddatait/data.html}{  \underline{Data} } 
$\blacktriangleright$ \href{http://homes.cs.washington.edu/~fire/reddatait/code.html}{  \underline{Code}} 
\\ \\
\textbf{Keywords:} Complex Network Models;  Online Social Networks; Reddit  
\end{abstract}
}

\begin{document}

\maketitle


\section{Introduction}

\lettrine[nindent=0em,lines=3]{C}omplex networks are loosely defined as networks with non-trivial structure and dynamics, 
appearing in social, molecular, biological, ecological, and economical real-world complex systems~\cite{estrada2013journal}.
Many studies have provided 
models and algorithms to explain how complex networks evolve~\cite{barabasi1999emergence,watts1998collective,holme2002growing,sallaberry2013model}. 
However, scores of questions remain unanswered~\cite{albert2002statistical}.

With the recent rapid advances in the field of data science, new algorithms, infrastructures, and techniques 
for data mining, data storage, data prediction, and data visualization have emerged~\cite{armbrust2010view,bostock2011d3,low2014graphlab,zaharia2010spark}.  
These tools make it feasible to gain new insights from large quantities of data. 

In this study, we utilize data science tools to thoroughly examine the evolution process of 
online social networks. Our study focuses on Reddit,\footnote{\url{http://www.reddit.com}} 
a huge collection of online communities that discuss what is new and popular on the Internet.
Namely, we utilize the recently published Reddit 
dataset, which contains over a terabyte of uncompressed data and
 consists of over 13.2 million users\footnote{The dataset contains 13,213,173 unique usernames. However, many Reddit users have several usernames.}
  who posted over 1.65 billion comments published over a 7-year time period. 
 This vast dataset contains information on the evolution process of more than 239,000 online communities and the 
connections among the community members (see Section~\ref{sec:dataset}). 
We used the Reddit dataset to construct a large corpus of 11,965 directed social networks, which
contain over 97\% of all the posted comments in the dataset (see Section~\ref{sec:dataset}). 
We assembled an unprecedented corpus of social networks with diverse topologies, ranging from
social networks with 11 vertices up to social networks with over 4 million vertices (see Table~\ref{tab:features}). 

\begin{figure*}
  \centering
  \includegraphics[width=0.7\textwidth]{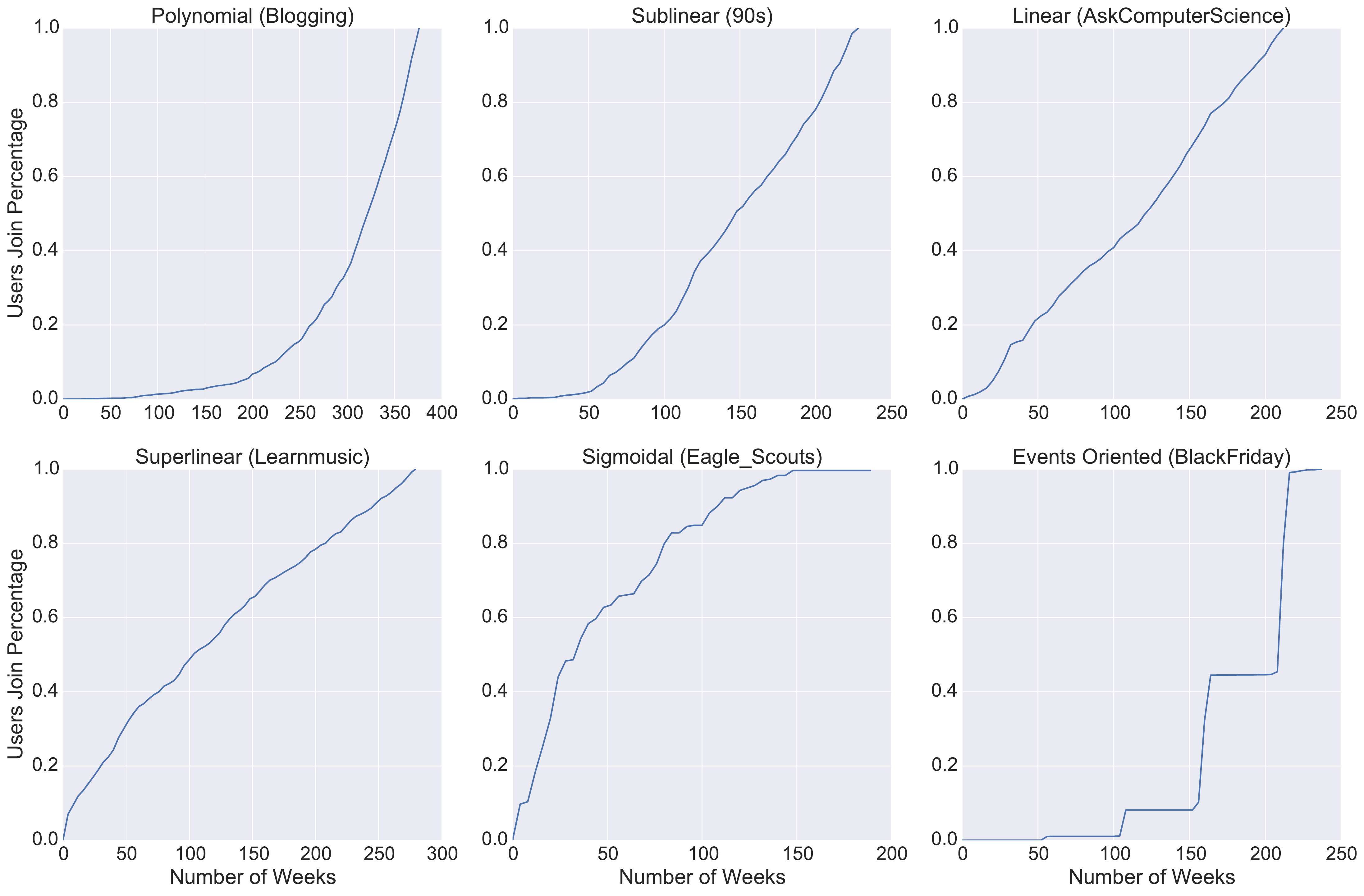}
  \caption{Common User-Join Curves of the following subreddits: Blogging, 90s, AskComputerScience, Learnmusic, Eagle-Scouts, and BlackFriday.}
      \label{fig:patterns}
\end{figure*}

By closely examining the social networks' evolution process, we observed that 
users tend to connect to those users having similar joining times. For example, we discovered that the median join-time difference between two connected users was 350.79 days.
 Moreover, we observed that the likelihood of a link being  
established between two users decreases sharply as the join-time difference increases (see Section~\ref{sec:timediff}).  
For instance, only 3.2\% of the examined edges were between two users with a join-time difference over 4 years (see Figure~\ref{fig:timedifference}). 
Furthermore, 
 we identified six common patterns in which users joined the network (see Section~\ref{sec:uac_category} and Figure~\ref{fig:patterns}).  
Moreover, we observed that different user-join patterns influence the 
structures of the social networks. For example, networks with sigmoidal-like growth usually have denser 
and higher average clustering coefficients than networks with polynomial growth (see Section~\ref{sec:ujc}).
The majority of random network generation models do not take into account either the 
time or the rate that new vertices join the network (see Section~\ref{sec:related} and Sallaberry et al.~\cite{sallaberry2013model}).

Inspired by the above observations, we developed the Temporal Preferential Attachment random network generation model (denoted TPA model). 
This model generalizes the well-known Barab\'{a}si-Albert network generation 
model (denoted BA model)~\cite{barabasi1999emergence} by adding to the model the rate in which vertices join the network, as well as 
storing each vertex's arrival time. Moreover, the model takes as input the 
probability that each vertex will connect to other vertices with the same or with a different 
join-time (see Section~\ref{sec:tpa}).  

We demonstrate that the presented TPA model is able to 
produce arbitrary-sized random scale-free networks with relative high clustering coefficients, which are sensitive  
to vertex arrival times and rates (see Section~\ref{sec:tpaeval}). This model mimics realistic social 
networks and should help to better understand the evolution process of complex 
networks.

The remainder of the paper is organized as follows: In Section~\ref{sec:related}, we provide an overview of various related studies. 
In Section~\ref{sec:dataset}, we give details on the Reddit dataset and present our methodology to 
clean the dataset.
Section~\ref{sec:sn} describes how we constructed the 
subreddit social networks and outlines the social network topological features we utilized in this study. 
Next, in Section~\ref{sec:ujc}, we analyze the networks' temporal dynamics.
Subsequently, in Section~\ref{sec:tpa}, we introduce the TPA model.
Then, in Section~\ref{sec:diss}, we discuss the obtained results. Lastly, in Section~\ref{sec:conclusions}, we present
our conclusions and also offer future research directions.

\section{Related Work}
\label{sec:related}

The study of complex networks began over half a century ago, in 1965. While studying a network of citations among scientific papers, Price observed  a network in which the degree distribution followed a power law~\cite{price1965statistical}. 
 Later, in 1976, Price
 provided an explanation of the creation of these types of networks: 
	 ``Success seems to breed success. A paper which has been cited many times is more likely to be cited again than one which has been little cited''~\cite{price1976general}. 
Price subsequently offered a method for the creation of networks in which the degree 
distribution follows a power law.

Several decades later, in 1998, Watts and Strogatz~\cite{watts1998collective} presented a model for generating 
small-world networks. Typically, small-world networks have a relatively high clustering coefficient, and the 
distance between any two vertices scales as the logarithm of the number of vertices~\cite{sallaberry2013model}. 
In the following year, Barab\'{a}si and Albert observed that degree distributions that follow power 
laws exist in a variety of networks, including the World Wide Web~\cite{barabasi1999emergence}.  Barab\'{a}si and Albert coined the 
term ``scale-free networks'' for describing such networks. 
Similar to Price's method ~\cite{price1976general},\footnote{A detailed explanation of the differences between Price's method 
and the Barab\'{a}si-Albert model for constructing complex networks can be found in~\cite{newman2003structure}.} 
Barab\'{a}si and Albert~\cite{barabasi1999emergence} suggested a simple and elegant model for creating random complex networks
based on the rule that the rich are getting richer. In the BA model, a network starts with $m$  
connected vertices. Each new vertex that is added (one at a time) has a greater 
probability of connecting to pre-existing vertices with higher degree, where the probability of connecting to an existing 
$v$ is proportional to $v$'s degree~\cite{barabasi1999emergence}. 
Consequently, ``rich'' vertices with high degrees tend to become even ``richer'' due to their 
connections with new vertices that join the graph. 

Many real-world complex networks have a community structure in which ``the 
division
of network nodes into groups within which the network connections are dense, 
but between which they are sparser''~\cite{newman2004finding}. In 2004, Newman and Girvan 
proposed a community detection algorithm and offered a simple method to create 
networks with community structure~\cite{newman2004finding}.

 Even though the models described above can explain some of the characteristics of real-world 
complex networks, the random networks created by these models were lacking in other 
properties that were observed in real-world complex networks.
Therefore, in recent years, other models have been suggested which have additional characteristics~\cite{barabasi1999emergence,holme2002growing,sallaberry2013model,watts1998collective}. 

A similar study to ours was conducted by 
Leskovec et al.~\cite{leskovec2008microscopic}. They performed edge-by-edge analysis of 
four large-scale networks - Flickr, Delicious, LinkedIn, and Yahoo Answers - with time spans ranging 
from four months to almost four years. By studying a wide variety of network formation strategies, they observed that
edge locality plays a critical role in the evolution of networks, and they offered a model which focused on microscopic vertex behavior. In their proposed model, vertices arrive at a pre-specified rate and choose their 
lifetimes. Afterwards, each vertex ``independently initiates edges according to a 
`gap'
 process, selecting a destination for each edge according to a simple triangle-closing model free of any 
parameters''~\cite{leskovec2008microscopic}. They showed that their model could closely mimic the 
macroscopic characteristics of real social networks. Additionally, 
Leskovec et al., similar to our study, observed the arrival patterns of various vertices.
Namely, they observed that (a) Flickr's network data has grown exponentially; (b) Delicious has grown slightly superlinearly; (c) LinkedIn has grown 
quadratically;
and (d) Yahoo Answers has grown sublinearly. Due to these observations, they concluded that vertex arrival functions needed to be part of 
their proposed model.
However, their study did not analyze the implications of using different 
arrival functions. 

In this study, we extend the BA model. Our presented TPA model is able to 
generate small-world, scale-free networks, with relatively high clustering-coefficient 
values. Moreover, 
our model takes into consideration the time and rate in which vertices join the network.

\section{The Reddit Dataset}
\label{sec:dataset}
Reddit is a news aggregation website and online social platform, which was launched in 2005 by Steve Huffman and Alexis Ohanian~\cite{bergstrom2011don}.
Reddit users (also known as ``redditors'') can submit content on the website, which is then commented upon, and upvoted or downvoted by other users in order
to increase or decrease the submission visibility. 
Redditors can also create their own subreddit on a topic of their choosing, make it public or private, and let other redditors join it.
This makes Reddit a collection of online communities, centered around a variety of topics 
such as books, gaming, science,
 and asking questions.\footnote{Each subreddit web page can be accessed at the following URL: \url{https://www.reddit.com/r/<Subreddit Name>/},
 by replacing \textit{<Subreddit Name>} with the subreddit's name. }
 
 In this study, we utilized the Reddit dataset which 
 was recently made public by Jason Michael Baumgartner (see Section~\ref{sec:data}). The Reddit dataset contains  
over 1.65 billion comments that were posted from October 2007 through May 2015.
These posts were created by 13,213,173 users, with unique usernames, in 239,772 different subreddits.
The dataset contains information on the exact time and date each comment was 
posted. Moreover, the dataset contains each comment's ID, as well as 
information on the user who posted it and the ID of the parent comment, i.e., the ID to which the current
comment replied.

 For this study, we cleaned the dataset by removing nonessential 
comments, specifically those that were marked as deleted and those that did not include
 the information of the user who posted them. Additionally, we removed posts by 
 users who with high probability were bots. Namely, we removed all the users who posted more than 100,000 comments each, and we removed 897 redditors whose comments appeared in the bots list published in the BotWatchman 
 subreddit.\footnote{We downloaded the bots list from the BotWatchman 
 subreddit \url{https://www.reddit.com/r/BotWatchman/} during December 2015. } 
 After the removal of these posts, we were left with over 1.42 billion comments.

\section{Subreddit Social Networks}
\label{sec:sn}
In the following subsections, we give a detailed description of the methods that we used to construct and analyze the subreddit social networks.
In Section~\ref{sec:sn_construct}, we introduce the methods we used to 
construct each subreddit's social network. Then, in 
Section~\ref{sec:sn-features}, we present the topological features we extracted 
from each network. Additionally, we provide a statistical overview of each 
topological feature across all social networks.

\subsection{Social Network Construction}
\label{sec:sn_construct}
To perform the analysis of the subreddits' underlying social networks, we first needed to 
 construct these social networks. However, many of the subreddits did not 
 contain enough users or were not active for a long enough time to extract meaningful insights on the user-join patterns. For example, the 
 median number of redditors in a subreddit was 3, while only 5\% of the subreddits consisted of 371 redditors
 or more. 
 Therefore, for all subreddits in the clean dataset of over 1.42 billion comments, we 
 selected only those subreddits that had at least 10 users, consisted of at least 1,000 comments, 
 and were active\footnote{Throughout this study, we considered the time in which a subreddit was active as the time 
 difference between the first and last comments published in the subreddit. } for at least 1 year.
 Out of all the subreddits, 11,965 subreddits with over 1.38 billion posts (referred to as \textit{selected subreddits}) fulfilled this criteria. 
 
 Next, for each selected subreddit, similar to the construction method used by Kairam et al.~\cite{kairam2012life}, we created 
 the subreddit's social network directed graph by connecting users who posted comments as replies to other posted 
 comments. Namely, for a subreddit, we define the subreddit's directed graph to be:
 $G := <V,E>$ where $V$ is the set of vertices, representing all the subreddit's users who posted at least a single comment in
 the subreddit, and $e := (u,v) \in E$ is the list of all edges between the subreddit's users $u \in V$ and $v \in V$.
 We define an edge between $u$ and $v$ to exist if there exists a comment on the subreddit posted by $u$ to which $v$ posted a reply on the same subreddit. 
 Lastly, we used the Powerlaw Python package~\cite{alstott2014powerlaw} and observed that most of the 
 social networks' vertex connection distributions matched
  power law distributions with various exponent values.
 It important to notice that the constructed directed graph also includes single vertices of redditors who 
 posted comments and did not receive any reply, as well as self-loop edges of 
 redditors who posted a comment and then posted a comment as a reply to their own 
 comment.

\subsection{Calculating Topological Features }
\label{sec:sn-features}
 For each selected subreddit constructed social network graph, $G:= <V,E>$, we calculated the following 
 topological features:
  \begin{compactitem}   
    \item \textit{Vertices number} - the number of vertices in the graph, defined as $|V|$. 
    
    \item \textit{Edges number} - the number of edges in the graph, defined as $|E|$. 
    
    \item \textit{Density} - the graph density, defined as $D := \frac{|E|}{|V|\cdot 
    (|V|-1)}$.
    
    \item \textit{Number of self-loops} - the number of self-loops  in the 
    graph, defined as $Loops :=|\{(v,v) \in E| v \in V \}|$.
    
    \item \textit{Number of triangles} - the number of triangles (denoted by $|T|$) in the 
    graph~\cite{schank2007algorithmic}.
    
    \item \textit{Average clustering coefficient} - the graph's average 
    clustering coefficient (denoted by \textit{CC})~\cite{saramaki2007generalizations}.
    
    \item \textit{Degree-based features} - for a vertex $v \in V$, we defined the in-degree, out-degree, and total-degree of $v$ to be:
    \begin{compactitem}      
    \item $d_{in}(v) := |\{u \in V | \exists (u,v) \in E \}|$
    \item  $d_{out}(v) := |\{u \in V | \exists (v,u) \in E \}|$
    \item $d_{tot}(v) := |\{u \in V | \exists (u,v) \in E \mbox{ or }$  $\exists (v,u) \in E \}|.$
    \end{compactitem}
    Using the vertex degree definitions, we can define the following four degree features:
        \begin{compactitem} 
     \item $Avg\mbox{-}deg := \frac{\sum_{v \in V} d_{tot}(v)}{|V|}$ 
     \item  $Avg\mbox{-}in\mbox{-}deg :=\frac{\sum_{v \in V} d_{in}(v)}{|V|}$
     \item $Max\mbox{-}in\mbox{-}deg := max_{v \in V}(d_{in}(v))$; 
     \item $Max\mbox{-}out\mbox{-}deg$ $:= max_{v \in V}(d_{out}(v))$, 
        \end{compactitem} 
     where the \textit{Max} function returns the maximum value in a set.    
    \item \textit{Connected components-based features} - we separated the graph 
    into a set of weakly connected components (denoted by \textit{WCC})~\cite{wcc}, in which 
    $WCC := \{H \leq G| H \mbox{ is subgraph of } G\}$ and $G = \sqcup_{H \in WCC}H$.
     Using the $WCC$, we can also define the graph's largest component (referred to as $LC$) $G_{LC} := <V_{LC},E_{LC}>$,
     where $G_{LC} \leq G$, and $\forall H:=<V',E'> \in WCC, |V_{LC}| \geq |V'|$. 
    Using the above definitions, we can define the following five features:
            \begin{compactitem}
     \item \textit{Number of connected components} - the number of weakly 
      connected components, defined as $|WCC|$
      \item \textit{Largest component vertices number} - the number of vertices in $LC$, defined as $|V_{LC}|$
      \item \textit{Largest component edges number} - the number of edges in $LC$, defined as $|E_{LC}|$
      \item \textit{Largest component ratio} - the ratio between the number of 
      users in the largest component and all users in the subreddit, 
      defined as  $LC\mbox{-}Ratio := \frac{|V_{LC}|}{|V|}$ 
      \item \textit{The number of single components} - the number of components 
  in $WCC$ that consist of only a single vertex (denoted by \textit{|Single|}), defined 
      as $|\{H:=<V'',E''> \in WCC||V''|=1\}|$.             
      \end{compactitem}
  \end{compactitem}
  
  Additionally, for $v \in V$, we defined $v_{join}$ as the time $v$ was first active in the network.
 We then used this feature to calculate  a \textit{Days} feature, defined as \[Days(G) := max_{v \in V} v_{join}-min_{v \in V} v_{join}.\] The \textit{Days} feature is the number of days that had passed between the post times of  the first and last comments. Moreover, for each edge in the complete Reddit social network with  
  links, we calculated each edge time difference by \[e_{timediff} = |u_{join} - v_{join}|, \forall e=(u,v) \in E\]
   
  Table~\ref{tab:features} presents an overview of the various calculated topology features.
  Notice that the full set of topological features of all 11,965 selected subreddits is available 
  online (see Section~\ref{sec:data}).

  \begin{table*}
  \centering
  \caption{Subreddits Social Network Features Overview}
       \scalebox{0.85}{\begin{tabular}{|c|c|c|c|c|c|} \hline
Feature &Min& Max& Median& Mean & Std\\ \hline
Avg-deg & 0.0 & 53.343 & 1.893 & 2.889 & 3.4 \\
Avg-in-deg & 0.0 & 26.672 & 0.947 & 1.444 & 1.7 \\
CC & 0.0 & 0.941 & 0.025 & 0.047 & 0.067 \\
D & 0.0 & 0.936 & 0.001 & 0.003 & 0.017 \\
Days & 365.69 & 2,785.48 & 1,271.79 & 1,312.68 & 570.68 \\
LC-Ratio & 0.002 & 1 & 0.416 & 0.389 & 0.189 \\
Loops & 0.0 & 73,791 & 7 & 88.058 & 931.786 \\
Max-in-deg & 0.0 & 24,469 & 39 & 151.107 & 466.91 \\
Max-out-deg & 0.0 & 24,113 & 40 & 162.092 & 484.83 \\
|E$_{LC}$| & 0.0 & 42,973,517 & 768 & 30,118.416 & 484,602.04 \\
|E| & 0.0 & 42,980,043 & 860 & 30,287.708 & 484,771.67 \\
|Single| & 0.0 & 1,643,749 & 626 & 4,037.163 & 31,026.64 \\
|T| & 0.0 & 73,148,002 & 92 & 70,057.296 & 1,289,299.62 \\
|V$_{LC}$| & 1.0 & 2,391,502 & 341 & 4,345.554 & 39,425.02 \\
|V| & 11.0 & 4,043,528 & 1,110 & 8,605.852 & 70,357.23 \\
|WCC| & 1.0 & 1,647,816 & 663 & 4,132.8 & 31,247.4 \\    
\hline\end{tabular}}
    \label{tab:features}
\end{table*}
  
\section{Analyzing Temporal Dynamics of Networks} 
\label{sec:ujc}
\subsection{Network User-Join Curves}
In the following subsections, we describe in detail the methods which were used 
to construct and analyze user-join curves (denoted UJC). In Section~\ref{sec:uac_construct}, 
we define the UJC function and explain how we constructed the selected subreddit UJCs.
Next, in Section~\ref{sec:reg_analysis}, we describe the process we utilized to 
match each UJC and its corresponding function. Afterwards, in 
Section~\ref{sec:uac_category},
we present the methods used to categorize the different UJCs. Lastly, 
Section~\ref{sec:uac-ml} gives details on the methods used to predict the UJC 
categories based on the subreddits' topologies.

\subsubsection{User-Join Curve Construction}
\label{sec:uac_construct}
For all the selected subreddits, we constructed the UJCs using 
the following methodology: First, for each subreddit $S$, using the cleaned Reddit dataset, we calculated the number of 
weeks (denoted as $t^{S}_{end}$) between the first comment and the last comment that were posted on the subreddit. Afterwards, we 
defined $Users\mbox{-}Number_{S}(t)$ for $t \in [0,t^{S}_{end}]$ to be the number 
of users who joined the subreddit in $t$ weeks since the first comments were posted on the subreddit.
We also defined the overall number of users who joined 
the subreddit after $t^{S}_{end}$ weeks to be $Total\mbox{-}Users_{S}$. 
Then, using the above definitions, we defined $UJC:[0,t^{S}_{end}]$ \\ $ \rightarrow [0,1]$
as $UJC_{S}(t) = \frac{Users\mbox{-}Number_{S}(t)}{Total\mbox{-}Users_{S}}$,
where $UJC_{S}(0)$ and $UJC_{S}(t^{S}_{end})$ are always equal to 0 and 
1, respectively. 
Lastly, to create the UJCs for a network $S$, we calculated 
the $UJC_{S}$ values in 4-week intervals.\footnote{In case $t^{S}_{end}$  did not divide evenly by 4, the time interval between
the next-to-last and last UJC values was less than 4 weeks.}  
  By using this interval, the number of samples of the UJCs for each subreddit ranged from
15 to 101, with a median value of 47.

\subsubsection{User-Join Curve Regression Analysis}
\label{sec:reg_analysis}
To better understand the 11,965 UJCs that we created, 
we utilized CurveExpert software~\cite{hyams2010curveexpert} to match 
several selected UJCs with their best-fit functions using regression 
analysis. 
In most cases, the best fit was a high-degree polynomial function. To 
avoid over-fitting, we used the python-fit package\footnote{\url{https://pypi.python.org/pypi/python-fit/1.0.0}}
to find the polynomial function that was a best-fit for the majority of UJCs and still
had a relatively low degree. We discovered that 11,273 (94.2\%) and 9,199 (76.9\%) of the 
UJCs matched a quartic function (q(x) := $a + bX + cX^2 + dX^3 + eX^4$) with $R^2 \geq 0.95$ and $R^2 \geq 0.99$, 
respectively. 

Additionally, we performed regression analyses of the 692 UJCs that did not match 
quartic functions. Out of these UJCs, we observed that 274 matched the MMF model~\cite{hyams2010curveexpert}, $\frac{ab+cx^d}{b+x^d}$,
with $R^2 \geq 0.95$. The other 418 UJCs (referred to as \textit{anomalous UJCs}) presented a wide range of patterns.

\subsubsection{User-Join Curve Categorization}
\label{sec:uac_category}
\begin{figure}
  \centering
  \includegraphics[width=0.4\textwidth]{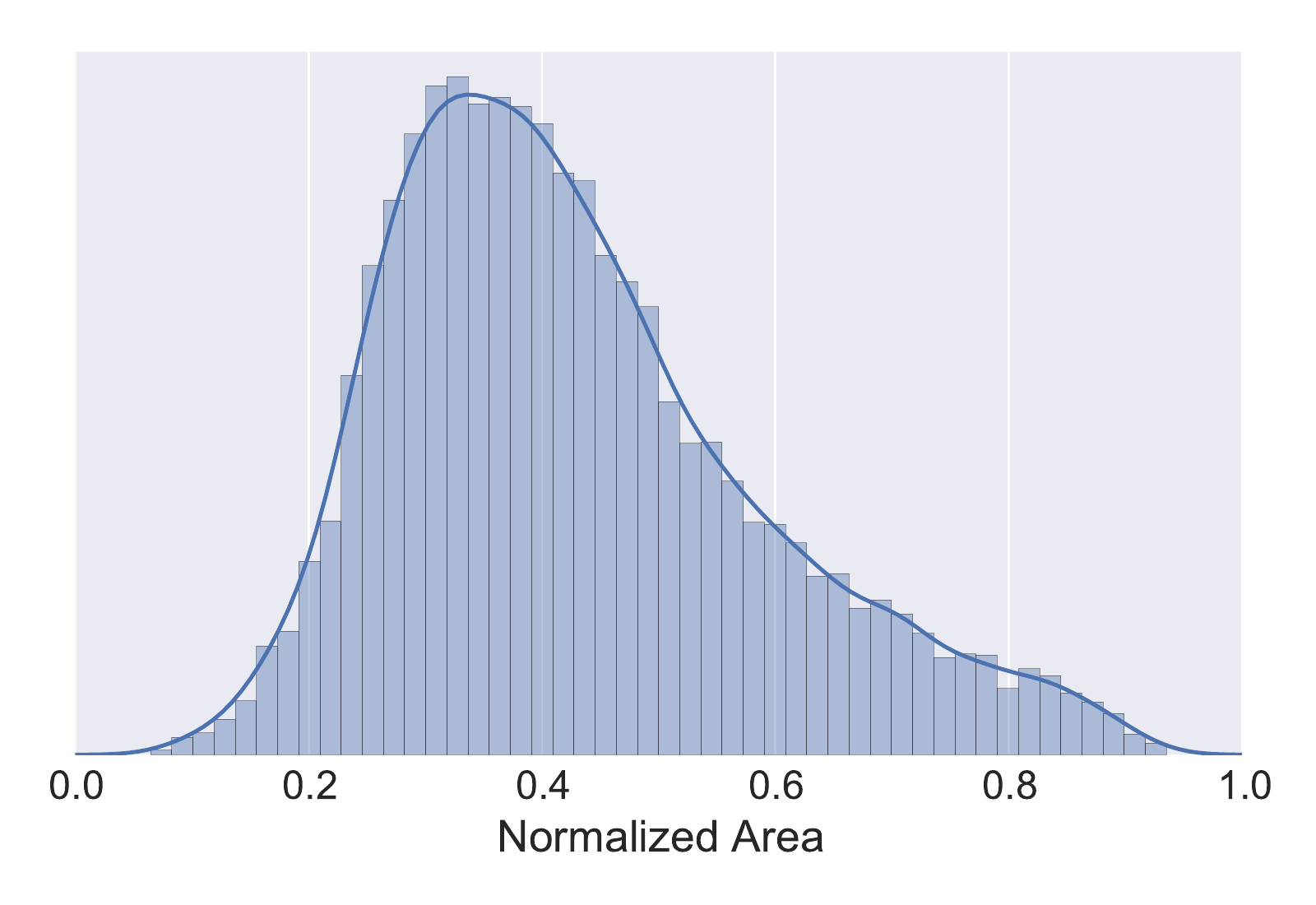}
  \caption{Normalized matched quartic functions' area distributions.}
      \label{fig:norm_area}
\end{figure}

After matching the UJCs with their best-fit quartic functions, we 
could now categorize the different UJCs.
For each matched quartic function $q_{S}(x)$ of subreddit $S$, we defined the 
normalized area function $ norm\mbox{-}area: q_{S} \rightarrow [0,1]$ as:
\[norm\mbox{-}area(q_S) := narea(q_S) := \frac{\int_{0}^{t^{S}_{end}} q_{S}(x) dx}{t^{S}_{end}}\]
In considering the 11,273 subreddit norm-areas, we observed that the 
norm-area distributions were skewed to the right, with a minimal value of 0.065, a maximal value of 0.935, and a median value of 0.4, with a standard deviation of 0.159 (see Figure~\ref{fig:norm_area}).

As a rule of thumb, we can foresee that fast growing subreddits, such as the Blogging subreddit (see Figure~\ref{fig:patterns}), will have a
relatively small norm-area value of close to 0. On the other hand, subreddits that had their growth 
halted for a relatively long time, such as the Eagle-Scouts subreddit (see Figure~\ref{fig:patterns}), will have a
norm-area of about 1, and subreddits with constant growth rates will have a norm-area 
of about 0.5.
Keeping this in mind, we divided the matched polynomials 
into five sets, according to their norm-area values: Set 1, $\{q_S| narea(q_S) \in [0,0.24) \}$ (with 905 UJCs); Set 2, 
$\{q_S| narea(q_S) \in [0.24,0.4) \}$ (with 4,733 UJCs);
Set 3, $\{q_S| narea(q_S) \in [0.4,0.56) \}$ (with 3,435 UJCs); Set 4, $\{q_S| narea(q_S) \in [0.56,0.72) \}$ (with 1,483 UJCs); and 
Set 5, $\{q_S| narea(q_S) \in [0.72, 1) \}$ (with 717 UJCs). For the five sets of 
matched polynomials, we calculated the various coefficient distributions and manually viewed the UJC graph visualizations for each 
set,\footnote{The UJC graphs are available for download; see Section~\ref{sec:data}.}
and we observed the following:
\begin{compactenum}
   \item The majority of the UJCs with $narea \in [0,0.24)$ (Set 1) have 
  positive $d$ and $e$ coefficients. Therefore, these UJCs typically have a cubic or \textit{quartic growth} rate (referred to as \textit{polynomial growth}).

  \item The majority of the UJCs with $narea \in [0.24,0.4)$ (Set 2) have a 
  negative $e$ coefficient. Additionally, by manually reviewing the UJCs, we  
  observed that in most cases these UJCs have a \textit{sublinear growth} rate. The growth typically starts relatively slowly and then changes into linear 
  growth.
  \item The majority of the UJCs with $narea \in [0.4,0.56)$ (Set 3) have 
    relatively large $c$ coefficients. However, in most cases the $d$ and $e$ coefficients have opposite signs; one is positive and the other is negative.
    This indicates that most of the UJCs do not grow at a fast rate. 
    By manually reviewing the UJCs in this set, we observed that these UJCs have a nearly \textit{linear growth} rate. 
  
  \item The majority of the UJCs with $narea \in [0.56,0.72)$ (Set 4) have a
  relatively high positive $b$ coefficient. However, in most cases the $d$ and $e$ coefficients have opposite 
  signs, indicating that most of the UJCs do not grow at a polynomial rate. Indeed, by manually examining
  the UJCs, we observed that most UJCs in this set have a \textit{superlinear growth} 
  rate. Many of the UJCs in this set started with a faster than linear growth 
  rate and then dropped to a linear growth rate.
  
  \item The majority of the UJCs with $narea \in [0.72,1)$ (Set 5) have 
  relatively high positive $b$ and $d$ coefficients, and low negative $c$ and $e$ 
  coefficients.
  This indicates that the UJCs grew very fast and then slowed down until the growth 
  stopped. Indeed, we observed a growth pattern in these UJCs which is similar to the \textit{sigmoidal growth} rate. 
\end{compactenum}

Additionally, we inspected the growth patterns of the 692 UJCs that did not match 
quartic functions. As mentioned in Section~\ref{sec:reg_analysis}, 274 of these UJCs matched the MMF model, which has a \textit{sigmoidal growth} rate.
A closer look at the remaining 418 anomalous UJCs revealed that 
external events often affected the growth, such as the 
launch of a new season of TV shows. 
We refer to this growth pattern as \textit{events-oriented growth}.



\subsubsection{User-Join Curve Prediction}
\label{sec:uac-ml}
One of this study's goals was to understand how various UJC growth 
rates affect networks' topologies. To achieve this goal, using the 
11,965 selected subreddits, 
we first calculated the Pearson correlations between the subreddit topological features (see Section~\ref{sec:sn-features})
and the subreddit normalized-area values (see Section~\ref{sec:uac_category}).
Then, we used regression algorithms to construct models which could predict a subreddit's 
normalized-area values. 
Lastly, we used different classification algorithms to construct classifiers that 
could predict UJC categories based on their topologies. 
In the rest of this subsection, we provide an overview of each method we used and the obtained results.

\textbf{Correlations.}
\label{sec:pearson}
We calculated the Pearson correlations between the 11,965 subreddit social network topological
features and their normalized areas. The obtained results indicate that there is a
weak positive correlation between norm-area and the network's average clustering coefficient ($r=0.23$),
and also between the norm-area and the network's density ($r=0.21$). Additionally, there is a negligible
negative correlation between the norm-area and the network's maximal in-degree ($r=-0.12$), as well as between the norm-area and the maximal out-degree ($r=-0.14$). Moreover,
there is a moderate negative correlation between the norm-area and the \textit{Days} 
feature ($r=-0.47$).


\textbf{Regression Analysis.}
\label{sec:reg_predict}
We constructed regression prediction models, which can predict a subreddit's UJC normalized 
area using GraphLab Create~\cite{low2014graphlab}. We evaluated three regression 
models: linear regression, Boosted Trees regression, and Random Forest 
regression. We created these models twice, one time using all 
the topological features described in Section~\ref{sec:sn-features} and the 
second time with all the features plus the $Days$ feature.
We evaluated the models using 10-fold cross validation and measuring the models' 
average value of (a) mean absolute error (\textit{MAE}), (b) mean square error (\textit{MSE}), and (c) root mean squared error 
(\textit{RMSE}).

Out of these regression models, the Boosted Trees regression 
presented the best results, which were slightly better than the linear and Random Forest models, with $MAE=0.125$, $MSE=0.027$, and $RMSE=0.163$,
using only the topological features; and $MAE=0.114$, $MSE=0.232$, and $RMSE=0.152$, using the 
topological features plus the $Days$ feature. Additionally, the linear regression model created using the
11,965 subreddits' topology features presented an $R^2$ value of 0.125, using only the topological features, and an $R^2$ value of 0.262, using all the topological features plus the $Days$ feature. 
 Moreover, in the linear regression model constructed with all the features,
 the $CC$, $D$, $|E_{LC}|$,$|V_{LC}|$, $Max\mbox{-}in\mbox{-}deg$, $Loops$, 
 $|T|$, and $|WCC|$ had positive coefficients. 
 The $Avg\mbox{-}deg$, $|E|$, $LC\mbox{-}Ratio$, $Max\mbox{-}out\mbox{-}deg$,
 $|Single|$, $|V|$, and  $Days$ had negative coefficients.

\textbf{Supervised Learning.}
We constructed supervised learning classifiers which can classify the category of 
the subreddit's UJC, based on only the 11,965 selected subreddits' topological features, in the 
following way: First, we created a labeled dataset with the six growth categories we defined in Section~\ref{sec:uac_category}.
Next, using the labeled dataset, we used WEKA~\cite{hall2009weka} and constructed various classifiers using 
the following algorithms: OneR, J48 decision tree, Logistic, K-Nearest-Neighbors (KNN) with $K=1$, Rotation Forest, and Random Forest. Then, we evaluated each classifier using the 10-fold cross validation method and calculated the 
classifier's AUC (area under the ROC curve) values. Lastly, we repeated the construction and evaluation process, only this time with only two growth categories -- polynomial growth and sigmoidal growth. 

Out of all the trained classifiers, the Logistic classifier obtained the best results, in terms of AUC, on both datasets. 
On the first dataset, with six categories, the Logistic classifier obtained the 
highest weighted average AUC and the highest correct classification percentage of 0.64 and 41.2\%, respectively.
 These results were considerably better than the simple OneR and KNN classifiers that
obtained weighted average AUCs of 0.52 and 0.53, and correct classification percentages of 37.4\% and 31.7\%, respectively. 
On the second dataset, which consisted of subreddits with only polynomial growth and sigmoidal growth, 
the Logistic classifier obtained the 
highest AUC and highest true positive rate of 0.82 and 0.81, respectively.
 These results were considerably better than the simple OneR and KNN classifiers that
obtained AUCs of 0.67 and 0.64, respectively.

%


\subsection{Vertices' Join-Time Difference}
\label{sec:timediff}
Similar to Price's observation~\cite{price1965statistical} that new papers tend to be cited more than older papers, we noticed that redditors tend to be more engaged with other redditors who joined the network at a similar time, and less engaged with redditors who became active either a long time before or after they did. To validate our observation, for each edge $e$ in 362,230,386 links in the full Reddit network, we calculated the join-time difference in days ($e_{timediff}$) between each edge's vertices (see Section~\ref{sec:sn-features}). Afterwards, we calculated the probability of each time difference value. Moreover, we utilized regression analysis to match a function that estimated the probability of two redditors connecting, based on the join-time difference. 

\begin{figure}

 \centering
  \includegraphics[scale=0.4]{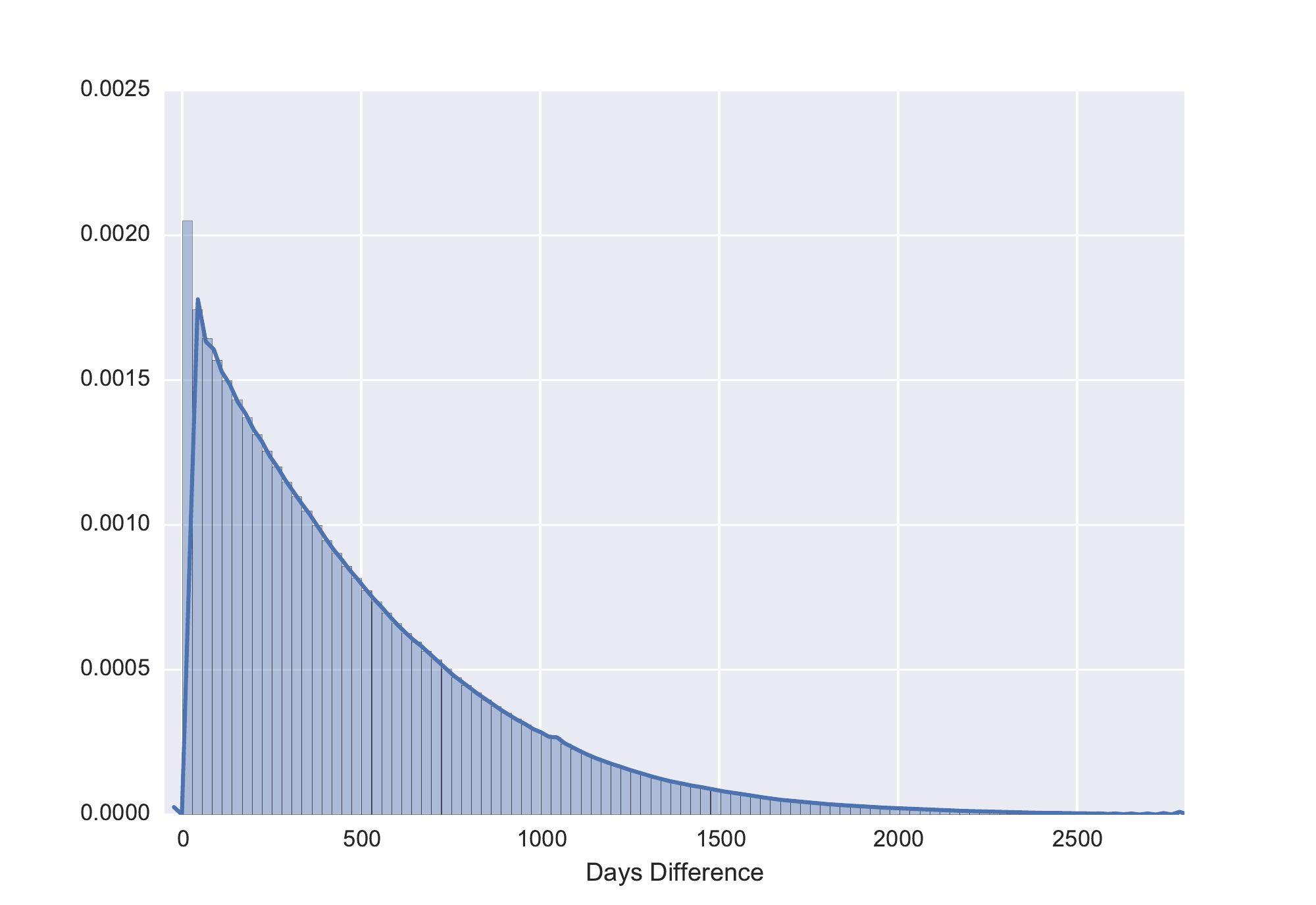}
 \caption{Edge time difference distribution.  It can be observed that as the join-time difference between two redditors increases, the probability of a link between them decreases sharply.  }
     \label{fig:timedifference}
 \end{figure}

The analysis results are presented in Figure~\ref{fig:timedifference}.  Additionally, using regression analysis, we discovered that 
the probability of two redditors with join-time difference $t$ to be linked can be estimated to be:
\[link\mbox{-}probability(t) \approx 0.002e^{-0.0019t}\] 

According to these results, we can observe that the probability of two redditors to be linked decreases sharply as the join-time difference between them increases. For example, while over  51\% of edges are between users with a join-time difference of less than a year, only 1\% of edges are between redditors with a join-time difference of over 5 years.

\section{The Temporal Preferential \\ Attachment Model}
\label{sec:tpa}
Inspired by the above results in which the vertex join-rate changes over time and edges tend to be formed among vertices with similar join times,
 we developed the TPA model for generating undirected random networks. The model is a generalization of the well-known BA model ~\cite{barabasi1999emergence}, with two main extensions: (a) instead of adding one vertex in each iteration to the network, the TPA model inserts a set of vertices in each iteration, where the set's size can be modified in each iteration; and (b) the probability of an edge forming between two vertices is negatively correlated with the time gap between the two vertices' join times. 

In the rest of this section, we first formally define the TPA model. Subsequently, we evaluate the model by using it to generate various random networks of different sizes, and we compare the constructed networks'  properties to the properties of real-world networks as well as to random networks created by other models. 

\subsection{The TPA Model Algorithm }
\begin{algorithm}
\footnotesize
  \caption{The Temporal Preferential Attachment Model Algorithm Overview}

   \begin{algorithmic}[1]

\Procedure{GenerateRandomGraph}{}
    \Require m, l, f 
     \State $g  \gets$ empty undirected graph
     \State $TimeGroupsList \gets$ empty list
	\For  {$i=0$ to $l.length$}     
		\State $NewVerticesList \gets$  list of $l[i]$ new vertices
		\State AddNewVerticesTimeGroup($NewVList$,$i$)
		\State AddVerticesToGraph(g $NewVList$)  

		\State AddToList(TimeGroupsList, $i$)
 		\For  {$v$ in $NewVerticesList$}
 			\State AddRandomEdges(g,v,m,f)
		\EndFor 
	\EndFor      
\EndProcedure
\\
\Procedure{AddRandomEdges}{}
    \Require g,v,m,f 
	\For  {$j=0$ to $m$}
	\Repeat
		\State $r \gets$ SelectFromList(TimeGroupsList,f)
		\State $u \gets$  GetVertexInGroupByDegree(g,r)
	\Until {$(v,u) \notin g$}
	\State AddEdge(g,v,u)

	\EndFor 
\EndProcedure
  \end{algorithmic}
\end{algorithm}

An overview of the TPA model algorithm is presented in Algorithm 1. 
The TPA model receives as input three parameters:
first, the number of edges (denoted $m$) to attach a new vertex to existing vertices;
second, an integers list (denoted $l$) with the number of vertices to add to the graph in each iteration; and third,
 a function (denoted $f: \mathbb{N} \to \mathbb{R} $) that given a time difference value,  returns the relative probability of an edge existing across two time groups. 
The algorithm starts by creating an empty undirected graph (line 1) and an empty time group list (line 2).
 Then, for each positive integer $l[i]$ in $l$, the algorithm does the following:
(1) creates new $l[i]$ vertices  with the time group set to $i$ (lines 5-6); (2) adds the new vertices to the graph (line 7); 
(3) adds $i$ to the $TimeGroupsList$ (line 8); and 
(4) connects each new added vertex to the other $m$ vertices using the AddRandomEdges procedure (line 10).

The AddRandomEdges procedure (lines 12-18) is the core of the model. The procedure receives as input five parameters: a graph ($g$), 
a vertex ($v$), the number of edges ($m$), a probability time difference function ($f$), and a list of existing time groups 
($TimeGroupsList$).
The AddRandomEdges procedure connects $v$ to $m$ other vertices in the graph using the following routine.
First, it randomly selects from $TimeGroupsList$ a time group (denoted $r$) where the probability of 
selecting each time group is given by $f$ (line 1).\footnote{Given $t_1,t_2,..,t_n$ 
time groups, the actual probability of an edge being created between two time groups with a time difference of  $d \leq n$ is equal to $\frac{f(d)}{\sum_{i \in[1,n]}{f(ti)}}$.  } 
 Then, similar to the BA model, the procedure selects one vertex ($u$) among all the vertices that are in the selected time group  $r$, 
 where vertices with higher degree have higher likelihood of being selected (line 16). In case the edge $(u,v)$ already exists in the graph, then the selection process of $u$ is repeated until a new $u$ in the graph is created.\footnote{In the Python implementation of the TPA model, to prevent cases where it is impossible to add new edges to $v$, we limited the number of repeats and moved on to connect the next vertex.  }
A detailed implemented TPA model in Python can be found in the paper's website (see Section~\ref{sec:data}).

To illustrate our TPA model algorithm, we can create a random graph using the following input parameters: $m=3 \mbox{, } l=[100,200,400]$, and $f(t)=2^{-1-t}$.
We start running the model with an empty graph. In the first iteration, we add 100 ($l[0]$) new vertices to the graph, and each new vertex has a  time group value of 0. In this iteration there are not any other time groups. Therefore, the 100 new vertices will only create 300 ($100\cdot 3$) edges among themselves in the following manner: each vertex will select 3 other vertices in the group, and similar to the BA model, vertices with higher degree will have higher probability of being selected, and the ``richer'' vertices will have a higher probability of becoming ``richer.''

In the second iteration, the model will insert 200 ($l[1]$) new vertices which will form 600 ($200\cdot 3$) new edges. However, this time we have two time groups: (a) a time group of 1 (with time difference 0), which contains all the 200 new vertices, and according to the time difference probability function, the probability of each new vertex establishing a connection to this group is
$f(0)=2^{-1-0} = 0.5$; and (b) a time group of 0 (with time difference of 1), which contains the previous 100 vertices, with a probability of $f(0)=2^{-1-1} = 0.25$ of connecting to vertices in this time group. According to these parameters, we can observe that the probability ratio of the two time groups is 2 to 1. Therefore, we can use this ratio to estimate that out of the 600 edges of the second iteration, about 400 edges will be formed among the vertices of time group 1,
 and about 200 edges will be formed among the vertices of time group 1 and time group 0, where each edge has a higher probability of connecting vertices with higher degree.
 
 Lastly, in the third iteration, our model will insert an additional 400 ($l[2]$) new vertices to the graph with a time group value of 2. These vertices will formulate 1200 ($400\cdot 3$) edges, of which about 686 will be among the vertices of time group 2; about 343 edges will be among the vertices of time groups 2 and 1 (time difference of 1), and about 171 edges will be among the vertices of time groups 2 and 0 (time difference of 2). Overall, the TPA model will have constructed a graph with 700 vertices and 2,100 edges.

\subsection{TPA Model Evaluation}
\label{sec:tpaeval}

To empirically evaluate the TPA model, we created various random networks by using various input parameters: 
\begin{compactenum}
\item The \textit{vertices number}  was set to three different sizes: 700, 6,200, and 12,350.
\item The \textit{edge number}, parameter $m$, was set to 3, creating networks with about 2,100, 18,600, and 37,050 edges.
\item We used linear, polynomial, and sigmoidal \textit{vertices growth rates}. For the linear growth rate, we added 10 new vertices in each iteration. For the polynomial growth rate, we used the sequence of 5, 20, 45, ..., $5\cdot x^2$, with a maximal $x$ value of 8, 16, and 20 for creating networks with 2,100, 18,600, 37,050 edges, respectively. For the sigmoidal growth rate, we used the same growth sequence as used in polynomial growth only in reverse order.
\item We used $f(t)=2^{-1-t}$  and $f(t)=0.8\cdot 0.2^t$ functions as \textit{time difference functions ($f$)}, where the $f(t)=0.8\cdot 0.2^t$ will create considerably more edges among all the vertices in the same time group than $f(t)=2^{-1-t}$.
\end{compactenum}

Overall, we assembled 18 different parameter settings for generating random networks. For each parameter setting, we utilized the TPA model to create 18 random networks. Subsequently, for each network, we calculated the network's average clustering coefficient (denoted CC), the maximal degree of vertex in the network (denoted $d_{max}$), the network's average shortest paths value (denoted Avg. SP), and 
the power law function ($k^{-\gamma}$) that matched the degree distribution of the network. To reduce variance of the calculated features, we repeated the network construction process and feature calculations 10 times for each parameter setting and calculated the average value of each feature. The results of these calculations are presented in Table~\ref{tab:tpa}.
Furthermore, for comparing the TPA model to other models, we used both the BA model and the Watts-Strogatz model to generate random networks of similar sizes (see Table~\ref{tab:tpa}).\footnote{The $p$ parameter in the Watts-Strogatz model was set to  0.1.}

  \begin{table}
  \centering
  \caption{Random Networks' Topological Properties}
       \scalebox{0.60}{
  \begin{tabular}{|c|c|c|c|c|c|c|c|c|} \hline
    Model &|V| &|E| & Rate & $f$ & CC & $d_{max}$ & Avg. SP & $k^{-\gamma}$  \\ \hline
TPA & 700 & 2100 & Linear & $2^{-1-t}$ & 0.063 & 28.4 & 4.23 & $k^{4.41}$ \\
 TPA & 700 & 2095 & Poly. & $2^{-1-t}$ & 0.03 & 37.2 & 3.7 & $k^{4.13}$ \\
 TPA & 700 & 2100 & Sig. & $2^{-1-t}$ & 0.053 & 58.1 & 3.73 & $k^{3.7}$ \\
 TPA & 700 & 2100 & Linear & $0.8\cdot0.2^{t}$ & 0.15 & 28.0 & 5.78 & $k^{3.81}$ \\
 TPA & 700 & 2095 & Poly. & $0.8\cdot0.2^{t}$ & 0.07 & 43.8 & 3.96 & $k^{3.53}$ \\
 TPA & 700 & 2100 & Sig. & $0.8\cdot0.2^{t}$ & 0.094 & 54.0 & 4.19 & $k^{3.32}$ \\
 TPA & 6200 & 18600 & Linear & $2^{-1-t}$ & 0.045 & 32.8 & 13.62 & $k^{6.95}$ \\
 TPA & 6200 & 18595 & Poly. & $2^{-1-t}$ & 0.007 & 61.1 & 4.89 & $k^{4.07}$ \\
 TPA & 6200 & 18600 & Sig. & $2^{-1-t}$ & 0.012 & 119.5 & 4.92 & $k^{3.62}$ \\
 TPA & 6200 & 18600 & Linear & $0.8\cdot0.2^{t}$ & 0.141 & 31.6 & 31.03 & $k^{3.18}$ \\
 TPA & 6200 & 18595 & Poly. & $0.8\cdot0.2^{t}$ & 0.025 & 97.4 & 5.59 & $k^{3.37}$ \\
 TPA & 6200 & 18600 & Sig. & $0.8\cdot0.2^{t}$ & 0.029 & 113.2 & 5.79 & $k^{3.26}$ \\
 TPA & 12350 & 37050 & Linear & $2^{-1-t}$ & 0.044 & 33.1 & 24.23 & $k^{7.14}$ \\
 TPA & 12350 & 37045 & Poly. & $2^{-1-t}$ & 0.004 & 81.7 & 5.3 & $k^{4.3}$ \\
 TPA & 12350 & 37050 & Sig. & $2^{-1-t}$ & 0.007 & 155.4 & 5.36 & $k^{3.72}$ \\
 TPA & 12350 & 37050 & Linear & $0.8\cdot0.2^{t}$ & 0.139 & 33.0 & 60.69 & $k^{7.15}$ \\
 TPA & 12350 & 37045 & Poly. & $0.8\cdot0.2^{t}$ & 0.017 & 116.0 & 6.27 & $k^{3.35}$ \\
 TPA & 12350 & 37050 & Sig. & $0.8\cdot0.2^{t}$ & 0.02 & 163.6 & 6.51 & $k^{3.34}$ \\
\hline
 BA & 700 & 2091 & - & - & 0.04 & 82.9 & 3.36 & $k^{3.13}$ \\
 BA & 6200 & 18591 & - & - & 0.008 & 234.4 & 4.12 & $k^{3.07}$ \\
 BA & 12350 & 37041 & - & - & 0.005 & 362.2 & 4.36 & $k^{3.04}$ \\
\hline
 WS & 700 & 2100 & - & - & 0.444 & 8.9 & 5.76 & - \\
 WS & 6200 & 18600 & - & - & 0.442 & 9.8 & 8.11 & - \\
 WS & 12350 & 37050 & - & - & 0.443 & 10.1 & 8.85 & - \\
 \hline
        \end{tabular}}
        \label{tab:tpa}
	\end{table}

\section{Discussion}
\label{sec:diss}
By analyzing the results presented in Sections~\ref{sec:sn},~\ref{sec:ujc}, and~\ref{sec:tpa}, the following can be noted:

First, from the construction process results presented in Section~\ref{sec:sn_construct}, 
we can observe that most of Reddit's online communities have an activity period of less than a 
year. Moreover, we can observe that the 11,965 selected subreddits, which 
are about 5\% of all the subreddits in the dataset, contain 1.38 billion posts; these are 97.7\% of all the posts in the constructed clean Reddit dataset. 
This indicates that most communities are short-lived and contain 
significantly less content than what can be found in other more active communities.
Therefore, studies aiming to analyze online communities need to 
carefully select the communities they choose to study.

Second, from Table~\ref{tab:features} we can observe that the Reddit dataset 
contains many types of communities with a wide range of size scales and with various 
topologies. However, from the results presented in Section~\ref{sec:reg_analysis}, we can surprisingly observe that even though there are many different social networks and the reasons people join these
communities are varied and complex, in most online communities the UJCs
match quartic functions. 

Third, according the correlation results presented in Section~\ref{sec:pearson}, 
we can observe that, in general, UJCs with higher normalized areas typically have higher CC values.
This observation aligns with the networks generated by the TPA model, in which networks created by sigmoidal growth usually presented higher CC values than same-size networks  created by polynomial growth (see Table~\ref{tab:tpa}).
 
Fourth, by observing the regression results in Section~\ref{sec:reg_predict}, 
which presented models with relatively low RMSE values, and the
classification results presented in Section~\ref{sec:uac-ml}, which presented promising results with AUC of 0.82, 
we can infer that analyzing a social network 
topology can, in many cases, predict the normalized area size of the UJC that 
created the network. Therefore, in many cases, we can predict the general pattern 
in which users joined the social network. Moreover, according to the 
classification results, the difference is noticeable between networks created by polynomial growth and networks created by sigmoidal growth.

Lastly, according to the results presented in Table~\ref{tab:tpa}, it can be noted that the TPA model generates scale-free networks with similar degree distribution to networks created by the BA model, and with small average shortest path values. Additionally, the TPA model can generate networks with much higher CC values than the BA model. However, in most cases networks created by the TPA model presented much lower $d_{max}$ values than networks of the same size created by the BA model. Nevertheless, in the BA model most vertices with high degree are likely vertices that join the network in the first iterations. By setting the TPA model's vertices join rate and time difference functions, it is possible to create vertices with relatively high degree that join the network in later iterations. This can more accurately mimic a network's evolution process and provide additional insight on how a newly added vertex suddenly becomes popular, such as when a post becomes viral in social networks.

\section{Conclusions}
\label{sec:conclusions}
Over a half century ago, Price estimated that half of papers' citations contained 
new papers while the other half were to ``immortal'' papers~\cite{price1965statistical}. 
In the last decades, much of the focus of network generation models was on  connecting new vertices in the graph to the ``immortal'' vertices. 
In this study, we demonstrate the importance of time as part of the network evolution process. 
We achieved this by analyzing a large-scale dataset containing over 1.65 billion comments.
 We analyzed 11,965 social networks created from the dataset and showed that the network UJCs have a high impact on the networks' structure (see Section~\ref{sec:ujc}). 
 Moreover, we showed that most edges are created between users who joined the network in a time difference of less than a year (see Section~\ref{sec:timediff}).

Taking into account these observations, we developed the TPA model, which incorporates both the rate and time that vertices join the network (see Section~\ref{sec:tpa}).
Moreover, we demonstrated that the TPA model can generate scale-free networks with relatively high CC values 
 and relatively small average shortest path values (see Table~\ref{tab:tpa}). Even though the TPA model is fairly simple, 
 we believe it presents a more accurate model of the evolution process of complex networks than other preferential attachment-based models.
Moreover,  the TPA model will help in better understanding the role of time in the network evolution process.

In the future, we hope to analytically analyze the TPA model and create a similar model for creating directed and weighted networks. 
Additionally, we plan to explore how different input parameters influence the created network topology. 
Another research direction we hope to pursue is to analyze the effect of different UJCs on the topologies of various complex networks, such as biological networks. 
 Moreover, the large corpus of social networks created and released as a result of this study can greatly contribute to better understanding 
 online communities and complex networks. 
 According Albert-L\'{a}szl\'{o} Barab\'{a}si, ``If data of similar detail 
capturing the dynamics of processes taking place on networks were to 
emerge in the coming years, our imagination will be the only limitation to 
progress''~\cite{barabasi2009scale}. 

\section{Data and Code Availability}
\label{sec:data}
Instructions on how to download the raw Reddit dataset are available in the 
following \href{https://www.reddit.com/r/datasets/comments/3bxlg7/i_have_every_publicly_available_reddit_comment}
{Reddit post}.
 Additionally, the dataset can be downloaded from this \href{http://files.pushshift.io/reddit/}{link}.
  This study is reproducible research. 
  Therefore, the social network datasets and a considerable part of the study's code, including implementation of the TPA model, are
   available at the project's \href{http://homes.cs.washington.edu/~fire/reddatait/}{website}, 
   which also gives researchers the ability to interactively explore and better understand the social networks in this study's dataset.

\section{Acknowledgments}
First and foremost, we would like to thank Jason Michael Baumgartner for 
making the Reddit dataset available online. We would like to thank Carol Teegarden for editing and proofreading this article to completion. 
We also want to thank Stephen Spencer for his IT expertise, which greatly assisted 
us during this research. Additionally, we thank the AWS Cloud Credits for Research. 
We also thank the Washington Research
Foundation Fund for Innovation in Data-Intensive Discovery, and the Moore/Sloan Data Science Environments Project at the University of
Washington for 
supporting this study.

\bibliographystyle{abbrv}
\bibliography{sigproc}  
\end{document}